\documentclass[conference]{IEEEtran}
\IEEEoverridecommandlockouts

\usepackage{cite}
\usepackage{amsmath,amssymb,amsfonts}
\usepackage{algorithm}
\usepackage{algpseudocode}
\usepackage{graphicx}
\usepackage{textcomp}
\usepackage{xcolor}
\usepackage{hyperref, tabularx, adjustbox, makecell, diagbox, multicol, multirow, xcolor, refcount, pifont}
\usepackage{booktabs}
\usepackage{orcidlink}

\newcommand{\cmark}{\textcolor{green!60!black}{\ding{51}}}
\newcommand{\xmark}{\textcolor{red!70!black}{\ding{55}}}

\def\BibTeX{{\rm B\kern-.05em{\sc i\kern-.025em b}\kern-.08em
    T\kern-.1667em\lower.7ex\hbox{E}\kern-.125emX}}
\begin{document}

\title{UniSAE: Unified Speech Attribute Editing on Speaker, Emotion and Low-Level Content via Discrete Phonetic Posteriorgram Modelling
\thanks{* Corresponding authors.}
}






\author{
    \IEEEauthorblockN{
    Chuanbo Zhu~\orcidlink{0009-0003-1023-0432}
    }
\IEEEauthorblockA{
    \textit{The Hong Kong University of}\\
    \textit{Science and Technology}\\
    Hong Kong SAR, China \\
    czhuat@connect.ust.hk
}

\and
    \IEEEauthorblockN{
    Wuyou Zhou~\orcidlink{0009-0003-5265-4321}
    }
\IEEEauthorblockA{
    \textit{The Hong Kong University of}\\
    \textit{Science and Technology}\\
    Hong Kong SAR, China \\
    wzhouba@connect.ust.hk
}

\and
\IEEEauthorblockN{
    Rongxiu Zhong
}
\IEEEauthorblockA{
    \textit{China Mobile}\\
    Beijing, China \\
    zhongrongxiu@\\cmjt.chinamobile.com
}

\and
\IEEEauthorblockN{
    Shilei Zhang
}
\IEEEauthorblockA{
    \textit{China Mobile}\\
    Beijing, China \\
    zhangshilei@\\cmjt.chinamobile.com
}

\and
\IEEEauthorblockN{
    Kun Qian ~\orcidlink{0000-0002-1918-6453}}
\IEEEauthorblockA{
    \textit{Beijing Institute of Technology}\\
    Beijing, China \\
    qian@bit.edu.cn}

\and
\IEEEauthorblockN{
    Yike Guo$^{*}$~\orcidlink{0009-0005-8401-282X}
    }
\IEEEauthorblockA{
    \textit{The Hong Kong University of}\\
    \textit{Science and Technology}\\
    Hong Kong SAR, China \\
    yikeguo@ust.hk
}

\and
\IEEEauthorblockN{ 
    Wei Xue$^{*}$~\orcidlink{0000-0002-4942-7748}}
\IEEEauthorblockA{
    \textit{The Hong Kong University of}\\
    \textit{Science and Technology}\\
    Hong Kong SAR, China \\
    weixue@ust.hk
}
}

\maketitle
\begin{abstract}
    Speech editing aims to modify specific portions of an utterance while preserving the remaining speech. Existing approaches primarily focus on word-level content modification and typically treat content, speaker, and emotion editing as separate tasks, limiting both editing granularity and flexibility.
    We propose UniSAE, a unified speech attribute editing framework which supports composable speaker, emotion and content editing from sub-phoneme to word level within a single architecture.
    UniSAE introduces a Discrete Phonetic PosteriorGram (DPPG) representation that factorizes speech content into discrete tokens encoding phoneme identity, pronunciation variants, and duration, enabling direct phoneme- and sub-phoneme-level editing.
    For higher-level modifications, an autoregressive content transformer predicts edited DPPG sequences for word-level content editing.
    The edited sequences are rendered into speech by a diffusion-based acoustic decoder, conditioned on disentangled speaker and emotion representations. Experimental results demonstrate that the proposed unified framework supports precise speaker and emotion control, content editing at multiple granularities, and joint modification of all three attributes within a single framework.
\end{abstract}

\section{Introduction}
\label{sec:intro}

Speech editing aims to modify an existing utterance while preserving the remaining speech~\cite{kassmann2024speech}. Recent advances in neural speech editing have enabled realistic content modification through generative models~\cite{peng-etal-2024-voicecraft, wang2025ssr}, supporting applications such as speech correction without re-recording entire utterances.
As these systems become increasingly capable, practical editing scenarios demand greater flexibility beyond simple word or phrase replacement.
For example, a sound director may wish to correct a pronunciation, adjust speaker characteristics, or modify emotional expression while maintaining consistency with the surrounding context. Such scenarios require control over multiple aspects of speech, spanning both linguistic content and paralinguistic attributes. These emerging requirements suggest that speech editing systems should support comprehensive and controllable manipulation of speech attributes rather than content alone.

\begin{table}[t]
\centering
\caption{
Comparison of editing capabilities across speech editing systems.
}
\footnotesize
\begin{adjustbox}{width=\columnwidth}
\label{tab:comparison}

\begin{tabular}{c c c c c c}
\toprule

Method
& Spk.
& Emo.
& Word
& Phoneme
& Sub-phoneme \\

\midrule

ZEST \cite{dutta2024ZEST}
& \cmark & \cmark & \xmark & \xmark & \xmark \\

EmoConv-Diff \cite{prabhu2024emoconv}
& \cmark & \cmark & \xmark & \xmark & \xmark \\

VoiceCraft \cite{peng-etal-2024-voicecraft}
& \xmark & \xmark & \cmark & $\triangle$ & \xmark \\

SSR-Speech \cite{wang2025ssr}
& \xmark & \xmark & \cmark & $\triangle$ & \xmark \\

\midrule

\textbf{UniSAE (Ours)}
& \textbf{\cmark}
& \textbf{\cmark}
& \textbf{\cmark}
& \textbf{\cmark}
& \textbf{\cmark} \\

\bottomrule
\end{tabular}
\end{adjustbox}

\vspace{0.5em}
{\footnotesize
\cmark: explicit support; \xmark: unsupported.\;
$\triangle$: achievable indirectly but without explicit control;\;

}

\end{table}

However, existing approaches largely treat content-based speech editing, voice conversion (VC)~\cite{yao2025stablevc}, and emotional voice conversion (EVC)~\cite{guo2023emodiff,dutta2024ZEST} as separate problems.
Meanwhile, recent advances in speech generation increasingly support unified control over speaker identity, emotion, and other speech attributes, enabling more flexible and expressive speech generation~\cite{Qwen3-TTS, zhou2025indextts2, chen2026actormind}. Extending similar controllability to speech editing would substantially broaden its practical applicability.
Therefore, we extend conventional speech editing to a broader task termed \textit{Speech Attribute Editing} (SAE), which treats linguistic content, speaker identity, and emotion as editable speech attributes within a unified framework. To make such a framework practically useful, several challenges should be addressed.

A primary challenge in SAE is enabling explicit and fine-grained editing of speech content.
Most previous works on speech editing rely on SSL or codec token representations that do not explicitly encode phonetic structure, leaving phoneme boundaries, pronunciation variants, and durations implicitly embedded within latent token sequences~\cite{peng-etal-2024-voicecraft, wang2025ssr}. As a result, phoneme-level modifications may be achieved implicitly through sequence generation, but explicit phonetic control and sub-phoneme editing, such as pronunciation variant and duration manipulation, remain largely unsupported.

Another major challenge is achieving reliable and independent control of multiple speech attributes.
Existing inpainting-based speech editing frameworks tightly couple content, speaker identity, and emotion within shared latent representations. Although such designs are effective at preserving source speech characteristics, they make independent manipulation of different attributes inherently difficult. Moreover, large-scale emotional speech corpora with diverse speakers and matched linguistic content are scarce, limiting the learning of robust and disentangled attribute representations. This challenge is further exacerbated by the intrinsic entanglement between speaker identity and emotion, which jointly influence prosody, spectral characteristics, and speaking style~\cite{ulgen2024revealing}. Consequently, modifying one attribute often induces unintended changes in the other, hindering reliable multi-attribute editing.

To address these challenges, we propose \textbf{UniSAE}, a unified framework for explicit speech attribute editing that jointly supports content, speaker, and emotion manipulation (Table~\ref{tab:comparison}).
For content editing, UniSAE introduces a Discrete Phonetic PosteriorGram (DPPG) representation that encodes phoneme, pronunciation-variant, and duration information, enabling fine-grained editing across multiple linguistic granularities.
To achieve independent control on multiple attributes, UniSAE adopts a two-stage architecture that disentangles content editing from acoustic attribute rendering through dedicated speaker and emotion representations. Furthermore, we construct UniEditCorpus via manifold distillation, a large-scale synthetic emotional speech corpus with counterfactual supervision that facilitates robust speaker--emotion disentanglement. Experimental results demonstrate that UniSAE achieves state-of-the-art speaker and emotion controllability while supporting word-, phoneme-, and sub-phoneme-level content editing within a unified framework. Audio samples are available at our demo page.\footnote{\url{https://anonymous260213.github.io/mydemo/}}

\begin{figure*}[t]
    \centering
    \includegraphics[width=\textwidth]{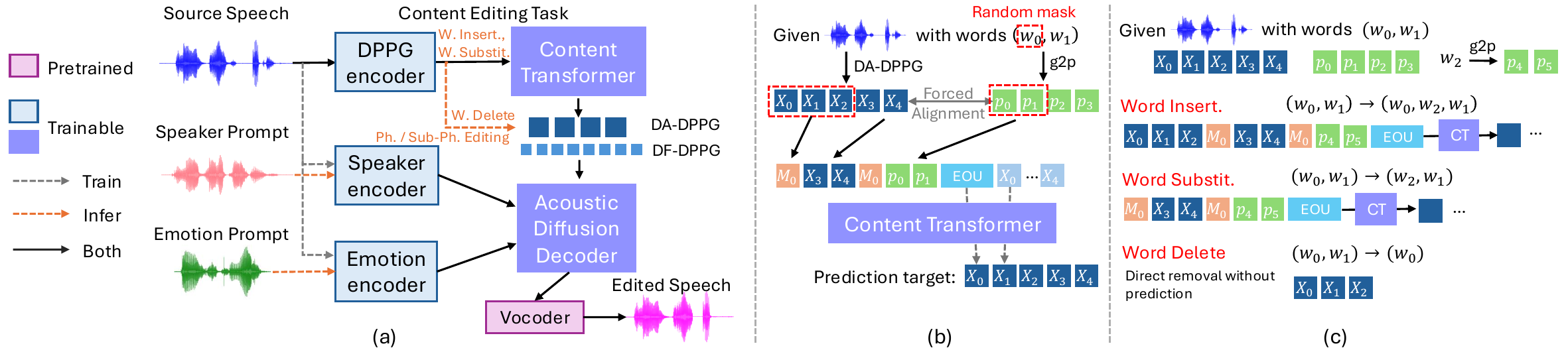}
    \caption{Overview of the proposed UniSAE framework, Content Transformer training and inference.}
    \label{fig:main}
\end{figure*}

\section{Unified Speech Attribute Editing: Problem Formulation}

\begin{table}[t]
\centering
\caption{
Taxonomy of speech attribute editing operations.
Each task is formulated as a state transition
$(c,s,e)\!\rightarrow\!(c',s',e')$, where $c$, $s$, and $e$
denote linguistic content, speaker identity, and emotion,
respectively. Joint editing is defined as the composition of
multiple primitive editing operations.
}
\label{tab:sae_task}

\small
\setlength{\tabcolsep}{8pt}

\begin{tabular}{ll}
\toprule
Task & Transition \\
\midrule

Speaker
& $(c,s,e)\!\rightarrow\!(c,s',e)$ \\

Emotion
& $(c,s,e)\!\rightarrow\!(c,s,e')$ \\

Word
& $(c,s,e)\!\rightarrow\!({c}'_w,s,e)$ \\

Phoneme
& $(c_{p, v, d},s,e)\!\rightarrow\!(c_{p',v, d},s,e)$\\

Sub-phoneme: variant
& $(c_{p, v, d},s,e)\!\rightarrow\!(c_{p,v', d},s,e)$ \\
Sub-phoneme: duration
& $(c_{p, v, d},s,e)\!\rightarrow\!(c_{p,v,d'},s,e)$\\

Joint & $(c,s,e)\!\rightarrow\!(c',s',e')$ \\

\bottomrule
\end{tabular}
\end{table}

We formulate speech attribute editing as controlled manipulation of three fundamental attributes of speech: linguistic content, speaker identity, and emotion. An utterance is represented as a triplet $(c,s,e)$, where $c$, $s$, and $e$ denote content, speaker identity, and emotion, respectively. Under this formulation, speech editing can be viewed as a state transition
$(c,s,e)\rightarrow(c',s',e'),$
where one or more attributes are modified while the remaining attributes are preserved.

To characterize the editing space, we define a set of primitive editing operations, summarized in Table~\ref{tab:sae_task}.
Speaker editing and emotion editing modify speaker identity and emotional attributes, respectively, while preserving the remaining factors. Content editing can be performed at multiple linguistic granularities. In addition to word-level editing, we introduce a finer phoneme-level representation by decomposing each phoneme into a triplet $(p,v,d)$ consisting of phoneme identity, phoneme pronunciation variant, and duration. This decomposition naturally enables phoneme editing through modification of $p$, as well as sub-phoneme editing through manipulation of $v$ or $d$ while preserving phoneme identity.
The primitive operations can be further composed, yielding a joint editing task of simultaneous modification of an arbitrary subset of speech attributes.

\section{UniEditCorpus: Manifold Distillation}

Learning disentangled speech representations is fundamentally limited by the scarcity of parallel emotional speech corpora, where speaker and emotion variations are sparsely observed and highly entangled. To address this limitation, we construct \textbf{UniEditCorpus}, a large-scale synthetic corpus with explicit control over content, speaker, and emotion through a process termed \textit{Manifold Distillation}. By providing abundant counterfactual supervision (e.g., identical content and speaker expressed with different emotions), UniEditCorpus serves as the primary training resource for UniSAE.

Specifically, we collect emotional speech prompts from six public emotional speech corpora and combine them with 2,000 utterances from VCTK~\cite{yamagishi2019vctk}. Using a zero-shot TTS model~\cite{deng2025indextts}, we synthesize all speaker--emotion combinations, resulting in a fully crossed content$\times$speaker$\times$emotion corpus containing 870k utterances (approximately 580 hours) from 87 speakers and five emotion categories.

To ensure data quality, synthesized samples are automatically filtered using pretrained speaker verification and emotion recognition models. The resulting corpus achieves a CER of 1.1\%, speaker consistency of 0.76, emotion consistency of 0.80, and an average UTMOS~\cite{saeki22c_interspeech} of 3.79, indicating high content fidelity and attribute consistency. Additional details on corpus construction and quality control are provided in the supplementary materials.

\section{UniSAE}
\label{sec: method}

UniSAE adopts a two-stage generation architecture consisting of content sequence modeling and acoustic rendering, illustrated in Figure~\ref{fig:main}(a).
Given source speech and an editing request, the model first produces an edited sequence of discrete phonetic tokens through the DPPG encoder and Content Transformer.
Then, an Acoustic Diffusion Decoder generates mel-spectrograms conditioned on the predicted content sequence together with disentangled speaker and emotion embeddings extracted by the corresponding encoders. The framework comprises four key components: (1) Discrete Phonetic PosteriorGram (DPPG) representation, (2) Content Transformer, (3) speaker and emotion encoders, and (4) Acoustic Diffusion Decoder.

\subsection{Discrete Phonetic Posteriorgram (DPPG)}
\label{subsec: DPPGs}
\begin{figure}
    \centering
    \includegraphics[width=0.75\linewidth]{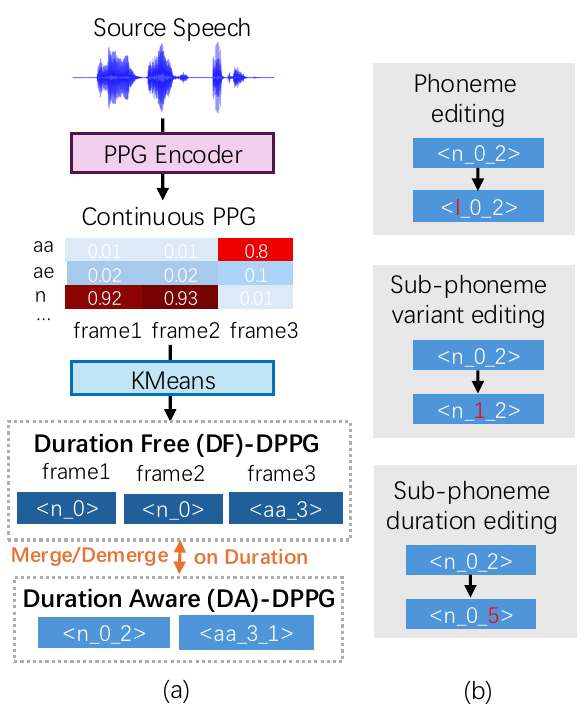}
    \caption{DPPG construction and phoneme, sub-phoneme editing operations.}
    \label{fig:dppg}
\end{figure}

The Discrete Phonetic Posteriorgram (DPPG) serves as the content representation in UniSAE. As illustrated in Fig.~\ref{fig:dppg}(a), source speech is first converted into continuous phonetic posteriorgrams (PPGs) using a pretrained PPG encoder and subsequently discretized via k-means clustering~\cite{lloyd1982kmeans}. The resulting discrete units preserve phonetic content while capturing pronunciation variations. Table~\ref{tab:dppg_example} shows example DPPG tokens for /n/, where each token captures a context-dependent phonetic variant, illustrating the expressiveness of DPPG.
Details of DPPG construction are provided in the supplementary materials.

To support both sequence modeling and acoustic generation, we employ two forms of DPPG: duration-free DPPG (DF-DPPG) \texttt{<p\_v>} and duration-aware DPPG (DA-DPPG) \texttt{<p\_v\_d>}, where $p$ denotes the phoneme identity, $v$ denotes a discrete phonetic variant, and $d$ denotes duration.
DF-DPPG is defined at the frame level, while DA-DPPG is obtained by merging consecutive identical DF-DPPG tokens and recording their duration. This factorized representation explicitly separates phoneme identity, pronunciation variation, and temporal realization while substantially reducing sequence length for efficient autoregressive modeling. The conversion between DF-DPPG and DA-DPPG is fully reversible through merge and demerge operations on the duration field.

As shown in Fig.~\ref{fig:dppg}(b), DPPG naturally supports editing at multiple linguistic granularities. Phoneme-level editing modifies the phoneme identity $p$, whereas sub-phoneme editing operates on the variant $v$ or duration $d$, enabling fine-grained control over pronunciation and timing.

\subsection{Content Transformer}

The Content Transformer is a GPT2-style decoder-only Transformer~\cite{radford2019language} that autoregressively models DA-DPPG tokens for word-level content editing as well as indirect phoneme modification. To support arbitrary span editing under left-to-right decoding, we propose a modified causal masking and rearrangement strategy inspired by VoiceCraft~\cite{peng-etal-2024-voicecraft}. Given a partially masked input sequence, the model generates the complete edited utterance by predicting DA-DPPG realizations from canonical phoneme sequences derived by the word transcript, jointly modeling phoneme identity, pronunciation variation, and duration.

\textbf{Training.} spans in the DA-DPPG sequence are masked using a mixture strategy: word-level masking with probability $\lambda$ (aligned to word boundaries) and token-level masking with probability $1-\lambda$ (random contiguous spans).
This mixed masking strategy enables the model to learn both word-level content editing and fine-grained phoneme-level editing within a unified framework.
The selected $N$ spans are replaced by mask tokens (\texttt{<M0>}...\texttt{<Mn>}), and their corresponding canonical phoneme sequences are appended to the input, preceded by mask identifiers. The correspondence between masked DA-DPPG spans and canonical phoneme sequences is determined using a DTW-based alignment algorithm. Details of the alignment procedure are provided in the supplementary materials.
For word-level masking, canonical phonemes are obtained via G2P~\cite{g2pE2019}; for token-level masking, the conditioning sequence retains only the phoneme identity, while variant and duration information are removed (\texttt{<p\_v\_d>} $\rightarrow$ \texttt{<p>}). An \textit{end of utterance} token \texttt{<EOU>} is appended to separate the conditioning sequence from the prediction target, after which the full target sequence is concatenated. 
The causal language modeling loss for training is defined as:
\begin{equation}
\mathcal{L}_{\text{CLM}} = - \sum_{t \in \mathcal{T}} \log P(Y_t \mid \mathbf{Y}_{<t})
\end{equation}
,
where $\mathcal{T}$ denotes the set of token positions after \texttt{<EOU>}. The model attends to the entire sequence, while the loss is computed only to tokens in $\mathcal{T}$. Figure \ref{fig:main}(b) shows an example of one word masking and the training procedure.

\textbf{Inference.}
For word-level insertion and substitution, the mask tokens are placed on the target positions, with the canonical phoneme sequences of the edited content appended after mask identifiers and before \texttt{<EOU>}. The model then generates the corresponding DA-DPPG tokens. For deletion, aligned DA-DPPG tokens are directly removed without prediction. Figure \ref{fig:main}(c) illustrates the examples of the three word editing tasks.

\begin{table}[t]
\centering
\caption{Discrete phonetic posteriorgram (DPPG) tokens of the phoneme /n/}
\label{tab:dppg_example}

\footnotesize
\setlength{\tabcolsep}{5pt}
\renewcommand{\arraystretch}{1.05}

\begin{tabular}{l l l}
\toprule
Token & Top-2 phonemes (probability) & Meaning \\
\midrule

\texttt{<n\_0>} & /n/ (98.9\%), /d/ (0.2\%) &  Canonical alveolar nasal\\
\texttt{<n\_1>} & /n/ (59.6\%), /ah/ (35.0\%) & Vowel-coarticulated variant\\
\texttt{<n\_2>} & /n/ (60.6\%), /d/ (32.8\%) & Voiced stop-influenced nasal\\
\texttt{<n\_3>} & /n/ (60.9\%), /sil/ (32.4\%) & Pause-adjacent nasal variant \\

\bottomrule
\end{tabular}
\vspace{-1em}
\end{table}

\subsection{Speaker and emotion disentanglement}
\label{subsec: disentangle}

A key advantage of UniEditCorpus is that each utterance is observed under multiple speaker--emotion combinations, providing explicit counterfactual supervision for disentanglement learning. To enable independent control of speaker identity and emotion during diffusion-based synthesis, we adapt the GE2E loss~\cite{ge2e} to a dual-attribute setting. Each training mini-batch contains $S\times E\times K$ utterances, where $K$ samples are drawn for every speaker--emotion pair. Speaker and emotion encoders are optimized separately using the same batch, allowing each encoder to cluster samples according to its target attribute while naturally marginalizing variations of the other attribute.

Given normalized embeddings $\mathbf{h}$ and class centroids $\mathbf{c}$, cosine similarity is used to measure embedding--centroid affinity. The loss for a target attribute (speaker or emotion) is defined as
\begin{equation}
    \mathcal{L}_{attr} = - \sum_{s,e,k} \log \frac{\exp(w \cdot \cos(\mathbf{h}, \mathbf{c}_{true}) + b)}{\sum_{j=1}^{N} \exp(w \cdot \cos(\mathbf{h}, \mathbf{c}_j) + b)},
\end{equation}
where $w$ and $b$ are learnable scaling factors and $N$ denotes the number of attribute classes. We optimize the speaker encoder using $\mathcal{L}{\text{spk}}$ and the emotion encoder using $\mathcal{L}{\text{emo}}$. This encourages attribute-discriminative embeddings while remaining invariant to the non-target attribute, providing disentangled conditioning signals for the diffusion decoder.

\subsection{Acoustic Diffusion Decoder}
\label{ssec:nar_diffusion}

The acoustic decoder is implemented as a diffusion probabilistic model (DPM) with velocity parameterization ($v$-prediction). 
Before decoding, DA-DPPG tokens \texttt{<p\_v\_d>} are expanded into frame-level DF-DPPG sequences.
The denoising network $f_\theta$ predicts the velocity $\mathbf{v}_t$ at diffusion timestep $t$, conditioned on DF-DPPG embeddings, speaker embeddings, and emotion embeddings. This disentangled conditioning enables independent control of speaker identity and emotional style while preserving phonetic consistency encoded in the DPPG sequence.
This design separates the the content editing from the acoustic rendering of paralinguistic editing.

\begin{table*}[!t]
\caption{Objective and subjective evaluation results of speaker and emotion editing on the UniEditCorpus and ESD test sets}
\label{result}
\begin{adjustbox}{width=\textwidth}
\begin{tabular}{l l ccc ccc ccc ccc}
\toprule
& & \multicolumn{6}{c}{UniEditCorpus} & \multicolumn{6}{c}{ESD} \\
\cmidrule(lr){3-8} \cmidrule(lr){9-14}

Setting & Method
& CER$\downarrow$ & SpkSim$\uparrow$ & EmoSim$\uparrow$ & nMOS$\uparrow$ & sMOS$\uparrow$ & eMOS$\uparrow$
& CER$\downarrow$ & SpkSim$\uparrow$ & EmoSim$\uparrow$ & nMOS$\uparrow$ & sMOS$\uparrow$ & eMOS$\uparrow$ \\
\midrule

\multirow{1}{*}{}
& GT
& 0.883 & 0.766 & 0.810 
& 4.158 & 4.368 & 4.474
& 2.433 & 0.751 & 0.941 
& 4.053 & 4.474 & 4.105 \\

\midrule
\multirow{3}{*}{Seen-Spk}

& EmoConv-Diff
& 4.900 & 0.707 & 0.721
& 3.895 & 3.000 & 2.842
& 10.000 & 0.683 & 0.625
& 3.105 & 2.789 & 2.316\\

& ZEST
& \textbf{2.119} & 0.555 & 0.758
& 2.579 & 2.263 & 3.684
& 9.012 & 0.544 & \textbf{0.768}
& 2.579 & 1.789 & 3.316 \\

& UniSAE
& 4.015 & \textbf{0.710} & \textbf{0.773}
& \textbf{4.053} & \textbf{3.474} & \textbf{3.947}
& \textbf{8.447} & \textbf{0.709} & 0.762
& \textbf{3.579} & \textbf{3.895} & \textbf{4.158} \\

\midrule
\multirow{4}{*}{Unseen-Spk}

& EmoConv-Diff
& 4.401 & 0.699 & 0.701
& 3.684 & 3.474 & 2.947
& 9.379 & \textbf{0.648} & 0.580
& 3.053 & \textbf{3.053} & 2.789 \\

& ZEST
& \textbf{1.959} & 0.560 & 0.753
& 2.579 & 2.263 & 3.053 
& 8.317 & 0.528 & 0.597
& 2.316 & 1.316 & 3.368 \\

& UniSAE
& 4.401 & \textbf{0.700} & \textbf{0.771}
& \textbf{4.105} & \textbf{3.684} & \textbf{4.421}
& \textbf{8.017} & 0.605 & \textbf{0.703}
& \textbf{3.684} & 2.105 & \textbf{3.579} \\

\bottomrule
\end{tabular}
\end{adjustbox}
\label{tab:decoder_evaluation}
\end{table*}

\section{Experiments}
\label{sec: experiment}

\textbf{Datasets.}
LibriTTS-R~\cite{libritts-r} and UniEditCorpus are used to train the DPPG tokenizer and Content Transformer, while UniEditCorpus additionally supervises the speaker/emotion encoders and Acoustic Diffusion Decoder. Evaluation is conducted on UniEditCorpus and ESD~\cite{zhou2021esd}, covering seen and unseen speakers as well as synthetic and human-acted emotional speech. The UniEditCorpus test set contains 10 seen and 5 unseen test speakers, while the ESD test set contains 5 seen and 2 unseen speakers across five emotions. Each test set includes 1,000 utterances, with 50 samples used for subjective evaluation. For word-level content editing, we construct ESDEdit, a benchmark derived from 100 ESD utterances by applying single-word insertion, deletion, and substitution operations.

\textbf{Tasks and baselines.}
We evaluate four tasks: speaker-emotion editing, word-level content editing, phoneme/sub-phoneme-level content editing and joint content-speaker-emotion editing. Since no prior framework supports all capabilities jointly, comparisons are task-specific. Speaker-emotion editing is compared with EmoConv-Diff~\cite{prabhu2024emoconv} and ZEST~\cite{dutta2024ZEST}; word-level editing with VoiceCraft~\cite{peng-etal-2024-voicecraft} and SSR-Speech~\cite{wang2025ssr}. Phoneme/sub-phoneme-level content editing and joint editing are evaluated only on UniSAE.

\textbf{Evaluation protocol.}
For speaker-emotion editing, given source speech $(c_A,s_B,e_C)$, a speaker prompt $(s_D,\mathrm{neutral})$, and an emotion prompt $(s_E,e_F)$, the target output is $(c_A,s_D,e_F)$.
Word-level content editing is evaluated on ESDEdit, with results averaged over insertion, deletion, and substitution operations. For joint editing, each ESDEdit utterance is additionally paired with speaker and emotion prompts from 2 seen speakers across 5 emotions.
Phoneme-level editing is evaluated on 8 phoneme substitution pairs (20 for each pair, 160 utterances in total), and results are reported as averages across all pairs; detailed pair definitions and per-pair results are provided in the supplementary material. Sub-phoneme editing replaces \texttt{<n\_0>} with \texttt{<n\_1>} on 20 utterances.

\textbf{Implementation details.}
K-means clustering on continuous PPGs produces 203 DF-DPPG tokens. The Content Transformer is a 12-layer decoder-only Transformer with a vocabulary of 5,440 DA-DPPG and phoneme tokens. Speaker and emotion encoders are 4-layer Transformers operating on wav2vec 2.0~\cite{baevski2020wav2vec} features and output 256-dimensional embeddings. The acoustic decoder is a U-Net diffusion model generating 24-kHz mel-spectrograms, followed by BigVGAN~\cite{lee2022bigvgan} vocoding. Additional details of the DPPG encoder are provided in the supplementary materials.

\textbf{Metrics.}
Content consistency is measured by CER using a pretrained ASR model~\cite{baevski2020wav2vec}. Speaker and emotion similarity (SpkSim, EmoSim) are computed using Resemblyzer~\cite{resemblyzer} and emotion2vec+~\cite{ma2023emotion2vec}, respectively. UTMOS~\cite{saeki22c_interspeech} is reported for word-level, joint-editing, and ablation experiments. For phoneme editing, a pretrained phoneme recognizer~\cite{churchwell2024ppg} is used to compute Target, Source, and Other Phoneme Detection rates (TPD/SPD/OPD), which sum to 100\%.

\textbf{Ablation settings.}
We study (1) replacing DPPG with continuous PPGs to assess the effect of discretization, and (2) replacing the disentangled speaker/emotion encoders with off-the-shelf pretrained d-vector~\cite{ge2e} and emotion2vec+~\cite{ma2023emotion2vec} representations.

\section{Results}
\label{sec: Results}

\begin{table}[t]
\caption{
Objective evaluation of word-level content editing on ESDEdit.
}
\label{tab:word_editing}
\centering
\begin{tabular}{lcccc}
\toprule
Method
& CER$\downarrow$
& SpkSim$\uparrow$
& EmoSim$\uparrow$
& UTMOS$\uparrow$\\
\midrule
VoiceCraft
& \textbf{6.556} & \textbf{0.894} & \textbf{0.933} & \textbf{3.615}\\

SSR-Speech
& 6.789 & 0.878 & 0.916 & 3.587\\

UniSAE
& 6.882 & 0.738 & 0.811 & 3.485 \\
\bottomrule
\end{tabular}
\end{table}

\begin{figure*}[t]
    \centering
    \includegraphics[width=\textwidth]{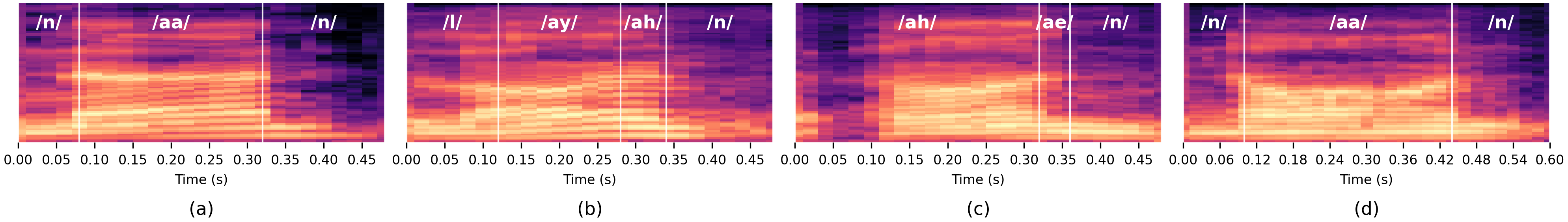}
    \caption{
    Mel spectrograms of the word ``nine'' under phoneme, sub-phoneme, and duration editing.
    (a) Original utterance with \texttt{<n\_0>};
    (b) phoneme editing (\texttt{<n\_0>} $\rightarrow$ \texttt{<l\_0>});
    (c) sub-phoneme editing (\texttt{<n\_0>} $\rightarrow$ \texttt{<n\_1>});
    (d) duration editing by lengthening /aa/ from 12 frames to 17 frames.
    }
    \label{fig:mel}
\end{figure*}

\subsection{Speaker and emotion editing}

We evaluate UniSAE on UniEditCorpus and ESD, with results summarized in Table~\ref{result}. The GT results show that UniEditCorpus achieves low CER and high similarity scores, approaching the quality of the human-acted ESD dataset despite being synthetically generated.

For content preservation of the editing results, UniSAE achieves CER comparable to the diffusion-based EmoConv-Diff on both datasets, while the HiFi-GAN-based ZEST attains the lowest CER. These results indicate that discretizing content representations into DPPG tokens introduces little information loss and preserves content accuracy during editing.

For speaker similarity, emotion similarity, and naturalness, UniSAE consistently outperforms the baseline systems on UniEditCorpus under both Seen-Spk and Unseen-Spk settings, indicating stronger disentanglement and more effective use of speaker and emotion prompts. On ESD, UniSAE maintains performance comparable to UniEditCorpus in the Seen-Spk setting, suggesting that the learned representations generalize across the synthetic-to-human-acted domain gap when speaker identities are observed during training. Performance degradation under the Unseen-Spk setting on ESD reveals remaining challenges in fully generalizing speaker representations to out-of-domain and unseen identities.

\subsection{Word-level content editing}

Table~\ref{tab:word_editing} presents word-level editing results on ESDEdit. VoiceCraft and SSR-Speech achieve lower CER and higher UTMOS, reflecting their specialization for word-level speech editing through codec-token inpainting and direct preservation of the original speech attributes. In contrast, UniSAE is designed as a unified framework that explicitly separates content, speaker, and emotion control. When the speaker and emotion prompts are set as source utterance, UniSAE can preserve the original speaker identity and emotional characteristics while performing content modification. Although its word-level editing performance is slightly lower than specialized inpainting-based systems, UniSAE additionally supports independent manipulation of speaker and emotion attributes as well as fine-grained phonetic editing. These results indicate that explicit phonetic modeling using DPPG combined with the Content Transformer, remains effective for word-level editing.

\begin{table}[t]
\centering
\caption{
Objective evaluation of phoneme-level content editing results averaged over eight phoneme pairs on ESD and a sub-phoneme variant editing case study.
}
\label{tab:phoneme_editing}
\small
\begin{tabular}{lcccc}
\toprule
Edit Type
& TPD
& SPD 
& OPD
\\
\midrule

Phoneme Identity Avg. (8 pairs)
& 83.75
& 7.50
& 8.75
\\

\midrule
\texttt{<n\_0>} $\rightarrow$ \texttt{<l\_0>} & 100 & 0 & 0 \\
\texttt{<n\_0>} $\rightarrow$ \texttt{<n\_1>}
& -
& 45 
& 55
\\

\bottomrule
\end{tabular}
\end{table}

\begin{table}[t]

\centering
\caption{
Objective evaluation of word-level content and speaker, emotion joint editing on ESDEdit.
}
\label{tab:joint_editing}
\resizebox{\columnwidth}{!}{%
\begin{tabular}{lcccc}
\toprule
Task
& CER$\downarrow$
& SpkSim$\uparrow$
& EmoSim$\uparrow$
& UTMOS$\uparrow$\\
\midrule
Cont. + Spk.
& 7.565 & 0.714 & 0.762 & 3.426\\
Cont. + Emo.
& 6.562 & 0.704 & 0.789 & 3.421 \\
Cont. + Spk. + Emo.
& 7.432 & 0.729 & 0.824 & 3.582 \\
\bottomrule
\end{tabular}}
\end{table}

\subsection{Phoneme- and sub-phoneme-level content editing}

As shown in Table~\ref{tab:phoneme_editing}, the high TPD and low SPD/OPD across eight phoneme substitution pairs demonstrate that DPPG enables reliable phoneme identity editing with limited unintended phoneme changes. The evaluated pairs cover diverse phoneme categories, including nasals, liquids, vowels, and fricatives, suggesting good generalization across phoneme types. Detailed results for individual phoneme pairs are provided in the supplementary material.

For sub-phoneme editing, we study phoneme /n/ in the word ``nine''. Unlike the phoneme substitution \texttt{<n\_0>} $\rightarrow$ \texttt{<l\_0>}, \texttt{<n\_1>} represents a vowel-coarticulated variant of the same phoneme. Replacing \texttt{<n\_0>} with \texttt{<n\_1>} reduces the detectability of the canonical /n/, with 55\% of outputs recognized as vowels, suggesting enhanced coarticulation while preserving phoneme identity (See Table~\ref{tab:dppg_example} for the definition of \texttt{<n\_0>} and \texttt{<n\_1>}).

Fig.~\ref{fig:mel} further visualizes the effects of phoneme-, variant-, and duration-level editing. Phoneme editing \texttt{<n\_0>} $\rightarrow$ \texttt{<l\_0>} changes the initial consonant from /n/ to /l/ and shifts the following vowel realization from /aa/ to a diphthong-like /ay ah/ sequence, indicating globally consistent pronunciation changes. Variant editing (\texttt{<n\_0>} $\rightarrow$ \texttt{<n\_1>}) weakens the nasal onset and advances the vowel transition, while duration editing lengthens /aa/ without affecting the surrounding phonetic content. These results demonstrate that UniSAE enables independent control of phoneme identity, pronunciation variants, and segment duration through discrete token manipulation.

\subsection{Joint speech attribute editing}

Beyond the speaker--emotion and content editing tasks evaluated above, we further investigate whether content editing can be composed with speaker and emotion control. 
Table~\ref{tab:joint_editing} shows that combining word-level content, speaker, and emotion editing causes little degradation compared with individual editing tasks. The stable CER, SpkSim, EmoSim, and UTMOS scores indicate effective disentanglement among the three attributes. Notably, the three-attribute editing setting $(c,s,e)\rightarrow(c',s',e')$ performs comparably to two-attribute editing, demonstrating reliable and composable control within a unified framework.


\subsection{Ablation study}

\begin{table}[t]
\caption{Ablation study on representation discretization and attribute disentanglement on UniEditCorpus seen-spk setting}
\centering
\resizebox{\columnwidth}{!}{%
\begin{tabular}{lcccc}
\toprule
Method
& CER $\downarrow$
& SpkSim $\uparrow$
& EmoSim $\uparrow$
& UTMOS $\uparrow$ \\
\midrule
UniSAE                  & 4.015 & 0.710  & 0.773  & 3.499 \\
D-PPG $\rightarrow$ PPG & 3.577 & 0.722  & 0.760  & 3.508  \\
Disent. Emb. $\rightarrow$ OTS Emb.   & 4.410 & 0.691  & 0.582  & 3.120  \\
\bottomrule
\end{tabular}
}
\label{tab:ablation}
\end{table}

Table~\ref{tab:ablation} presents ablation results on representation discretization and attribute disentanglement. Replacing DPPG with continuous PPGs yields comparable performance, indicating limited information loss from discretization. In contrast, replacing the proposed disentangled embeddings (Disent. Emb.) with off-the-shelf representations (OTS Emb.) substantially degrades speaker/emotion similarity and naturalness, suggesting that residual speaker-emotion entanglement in the pretrained representations encourages the model to rely on shortcut cues rather than the intended attribute conditioning.


\section{Conclusion}


We presented UniSAE, a unified framework that extends speech editing from word-level content replacement to general speech attribute editing. By combining explicit phonetic modeling with disentangled speaker and emotion control, UniSAE supports flexible editing of content, speaker identity, and emotion. Experimental results demonstrate effective and composable control across all three attributes while enabling fine-grained phonetic manipulation.

\section*{AI-Generated Content Disclosure}
\label{sec: genAI}

The authors disclose the use of generative AI tools in both the preparation and execution of this research. AI was utilized to polish the manuscript for clarity and to optimize experimental code. Additionally, this work incorporates synthetic data for model training. The effectiveness and reliability of this data are quantitatively demonstrated in the Results section. The authors have independently verified all AI-assisted outputs and data, remain fully accountable for the research findings, and consent to this submission.

\bibliographystyle{IEEEtran}
\bibliography{refs}

\clearpage
\renewcommand{\appendixname}{Supplementary Material}
\appendix

\section{Additional UniEditCorpus Details}
\label{Additional UniEditCorpus Details}

\subsection{Prompt pool}
Table~\ref{tab:dataset prompt} summarizes the emotional speech corpora used to construct the prompt pool for UniEditCorpus. All selected corpora contain recordings covering the five target emotion categories, namely neutral, happy, sad, angry, and surprised. We choose these datasets because they provide high-quality emotional speech recordings from diverse speakers while maintaining consistent emotion coverage across corpora. For datasets that include emotion intensity annotations, such as RAVDESS~\cite{livingstone2018ravdess} and MEAD~\cite{kaisiyuan2020mead}, only utterances labeled with strong emotional intensity are retained to ensure clear and unambiguous emotional expressions during synthesis. The final prompt pool consists of 87 speakers (46 male and 41 female), resulting in 870,000 synthesized utterances after exhaustive content--speaker--emotion combination.

 \begin{table}[htbp]
 \centering
 \caption {Prompt datasets used for UniEditCorpus, with number of speakers and resulting synthesized utterances.} 
 \label{tab:dataset prompt} 
 \begin{tabular}{llll} \hline
 Dataset  & \#Speakers & \#Emotions & \#Synthesis utterances \\ \hline
 ESD      & 10 (M: 6,  F: 4) &  5       & 100,000                 \\
 JLcorpus & 4   (M: 2,  F: 2) &  5       & 40,000                  \\
 RAVDESS  & 24 (M: 12, F: 12) & 5       & 240,000                 \\
 MEAD     & 46  (M: 26, F: 20) & 5     & 460,000                 \\
 TESS     & 2   (M: 0,  F: 2)  & 5      & 20,000                  \\
 EMNS     & 1   (M: 0,  F: 1)  & 5       & 10,000                  \\ \hline
 total    & 87 (M: 46, F: 41) & 5     & 870,000               \\  \hline
 \end{tabular}
 \end{table}

\subsection{Data partition}

Table~\ref{tab:dataset split} presents the data partition of UniEditCorpus. The corpus contains 2,000 unique textual contents, which are divided into 1,800, 100, and 100 contents for the training, validation, and test sets, respectively. To evaluate the generalization capability of UniSAE to unseen speaker identities, five speakers are held out from training and appear only in the validation and test sets. Consequently, the training set contains 77 speakers, while the validation and test sets each contain 82 speakers, including 77 seen speakers and 5 unseen speakers. The resulting corpus comprises 580.9 hours of speech and 870,000 utterances in total.

 \begin{table}[htbp]
\centering
\caption{Data split statistics of UniEditCorpus. Five speakers are held out during training and used only for validation and test to evaluate generalization to unseen speaker identities.}
\label{tab:dataset split} 
\resizebox{\columnwidth}{!}{
\begin{tabular}{lllll} \hline
Subset    & Duration & \#Utterances & \#Speakers &\#Unq. Contents \\ \hline
train     & 461h 39m & 693,000      & 77  &1800\\
valid     & 58h 10m  & 88,500       & 82 (77 seen, 5 unseen) &100\\
test      & 61h 2m   & 88,500       & 82 (77 seen, 5 unseen) &100\\ \hline
total     & 580h 51m & 870,000      & 87   &2000\\ \hline                         
\end{tabular}
}
\end{table}

\subsection{Quality Control and Corpus Validation}

To improve synthesis reliability, generated samples are filtered using a pretrained emotion recognition model. For each content--speaker--emotion combination, speech is regenerated until the predicted emotion matches the target label, with a maximum of 50 attempts. If all attempts fail, the sample with the highest target-emotion confidence is retained.

Corpus quality is evaluated using the same metrics as in the main paper, including CER, SpkSim, EmoSim, and UTMOS. The resulting corpus achieves a CER of 1.16\%, SpkSim of 0.759, EmoSim of 0.802, and UTMOS of 3.785. Since the UniEditCorpus test set used in the main paper is randomly sampled from the full corpus, its evaluation is representative of the overall dataset quality. The comparable performance between UniEditCorpus and ESD indicates that the synthesized corpus attains a quality level similar to that of a widely used human-recorded emotional speech dataset.

\section{Additional DPPG Implementation and Result Details}

\subsection{DPPG Implementation Details}

To construct the DPPG vocabulary, we discretize continuous phonetic posteriorgrams (PPGs) on a per-phoneme basis. Frame-level PPGs are extracted using a pretrained PPG encoder~\cite{churchwell2024ppg}, where each frame is represented as a posterior distribution over the 40 phoneme categories defined in the CMU Pronouncing Dictionary (CMUDict)\footnote{\url{http://www.speech.cs.cmu.edu/cgi-bin/cmudict}}. Training samples are collected from both LibriTTS-R~\cite{libritts-r} and UniEditCorpus with a sampling ratio of 0.4 and 0.6, respectively.

For each phoneme category, we gather all PPG frames whose highest posterior corresponds to that phoneme and retain at most 40,000 samples to ensure computational efficiency. MiniBatch K-Means clustering is then performed independently for each phoneme. The number of clusters is determined automatically using the elbow criterion with the Kneedle algorithm. Specifically, clustering minimizes the within-cluster variance

\[
I(k)=\sum_{i=1}^{n}\min_{\mu_j\in\mathcal{C}}|x_i-\mu_j|_2^2,
\]

where $x_i$ denotes a PPG frame and $\mu_j$ denotes a cluster centroid.

The resulting centroids define the sub-phoneme variants in DPPG. Given a PPG frame $x$ associated with phoneme $p$, its variant index is obtained through nearest-centroid assignment,

\[
v=\arg\min_j |x-\mu_j^{(p)}|_2^2.
\]

Each frame is then discretized into a token \texttt{<p\_v>}, where $p$ denotes the phoneme identity and $v$ denotes the corresponding sub-phoneme variant.

\subsection{Final DPPG Vocabulary Statistics}

Table~\ref{tab:dppg_all_clusters} reports the final number of K clusters of sub-phoneme variants automatically determined for all phoneme categories. As expected, different phonemes exhibit substantially different numbers of variants, reflecting their intrinsic acoustic variability. These results demonstrate that the proposed phoneme-wise clustering strategy adaptively allocates representational capacity according to the complexity of each phoneme, yielding a compact yet expressive discrete content vocabulary.

\begin{table}[t]
\centering
\caption{Optimal numbers of sub-phoneme variants automatically determined for all phoneme categories.}
\label{tab:dppg_all_clusters}
\small
\begin{tabular}{cccccccc}
\toprule
Phoneme & K & Phoneme & K & Phoneme & K & Phoneme & K \\
\midrule
/aa/ & 7 & /ae/ & 2 & /ah/ & 2 & /ao/ & 4 \\
/aw/ & 2 & /ay/  & 7 & /b/ & 7 & /ch/ & 7 \\
/d/ & 7 & /dh/ & 5 & /eh/ & 2 & /er/ & 7 \\
/ey/ & 6 & /f/  & 7 & /g/  & 7 & /hh/  & 8 \\
/ih/ & 2 & /iy/ & 7 & /jh/  & 2 & /k/  & 2 \\
/l/ & 2 & /m/ & 6 & /n/ & 8 & /ng/ & 5\\
/ow/ & 2 & /oy/ & 4 & /p/ & 5 & /r/ & 2\\
/s/ & 6 & /sh/ & 6 & /t/ & 6 & /th/ & 6\\
/uh/ & 5 & /uw/ & 5 & /v/ & 6 & /w/ & 6\\
/y/ & 7 & /z/ & 6 & /zh/ & 6 & sil & 3\\
\bottomrule
\end{tabular}
\end{table}

\subsection{DTW-based Phoneme-to-DPPG Forced Alignment}

To construct the conditioning sequence for word-level masked DA-DPPG prediction, we propose a lightweight phoneme-to-DPPG forced alignment algorithm that maps canonical phoneme sequences to DA-DPPG token spans.It combines monotonic dynamic time warping (DTW) with a phoneme-aware matching cost, enabling robust alignment despite pronunciation variations commonly observed in natural speech.

Given an input text, canonical phoneme sequences are first generated using a grapheme-to-phoneme (G2P) converter. Each DA-DPPG token of the form \texttt{<p\_v\_d>} is then reduced to its phoneme identity \texttt{} by removing the variant and duration fields, producing a phoneme sequence extracted from the acoustic representation.

Direct phoneme matching is often unreliable because canonical pronunciations do not always coincide with the phoneme identities observed in speech. Coarticulation, assimilation, and pronunciation variability frequently result in acoustically similar phoneme substitutions. To improve alignment robustness, the alignment algorithm employs a phoneme-aware matching cost (PAMC) defined as

\[
d_{\mathrm{pamc}}(p,q)=
\begin{cases}
0, & p=q,\\
0.25, & p,q \in \mathcal{G}_k,\\
0.5, & p,q \in \mathcal{V},\\
0.75, & p,q \in \mathcal{C},\\
1.0, & p\in\mathcal{V},q\in\mathcal{C}\ \text{or vice versa},\\
1.5, & p=\texttt{sil}\ \text{or}\ q=\texttt{sil},
\end{cases}
\]

where $\mathcal{G}_k$ denotes a fine-grained phonetic group, $\mathcal{V}$ denotes the set of vowels, and $\mathcal{C}$ denotes the set of consonants. Fine-grained phonetic groups are defined according to articulatory similarity, including vowel groups (e.g., {/iy/, /ih/}), stop groups (e.g., {/t/, /d/}), fricative groups (e.g., {/s/, /z/}), and sonorant groups (e.g., {/n/, /ng/}), as shown in Table~\ref{tab:phoneme_groups}. Consequently, substitutions between phonetically related phonemes incur substantially smaller penalties than substitutions between unrelated phonemes.

\begin{table}[t]
\centering
\caption{Phonetic groups used by PAMC}
\label{tab:phoneme_groups}
\footnotesize
\begin{tabular}{ll}
\toprule
Category & Groups \\
\midrule
Vowels & \makecell[l]{
\{/iy/, /ih/\},
\{/eh/, /ae/\},
\{/aa/, /ah/, /ao/\},\\
\{/uw/, /uh/\},
\{/ow/\},
\{/ay/, /aw/, /oy/, /ey/\},
\{/er/\}} \\

Stops &
\{/p/, /b/\},
\{/t/, /d/\},
\{/k/, /g/\} \\

Fricatives & \makecell[l]{
\{/f/, /v/\},
\{/th/, /dh/\},
\{/s/, /z/\},
\{/sh/, /zh/\},\\
\{/ch/, /jh/\},
\{/hh/\}} \\

Sonorants &
\{/m/\},
\{/n/, /ng/\},
\{/l/, /r/\},
\{/w/, /y/\} \\
\bottomrule
\end{tabular}
\end{table}

Using PAMC as the local matching cost, DTW is performed between the canonical phoneme sequence and the phoneme sequence extracted from DA-DPPG tokens. The alignment is constrained to be monotonic and forced: every canonical phoneme must align to at least one DA-DPPG token, while multiple consecutive DA-DPPG tokens may correspond to the same canonical phoneme. This formulation naturally accommodates duration variation while preserving phoneme order.

After DTW backtracking, the ending DA-DPPG token associated with each canonical phoneme is recovered. Aligned token spans are subsequently aggregated according to word boundaries, yielding a word-to-DPPG-span mapping. This mapping is then used to construct the conditioning sequence for word-level masked prediction. The complete procedure is summarized in Algorithm~\ref{alg:phonealign}.

\begin{algorithm}[t]
\caption{Phoneme-to-DPPG Alignment Algorithm}
\label{alg:phonealign}
\begin{algorithmic}[1]
\Require Text sequence $X$, DA-DPPG token sequence $Y=\{y_j\}_{j=1}^{M}$
\Ensure Word-to-DPPG-span mapping $\mathcal{A}$

\State Convert $X$ into word-level phoneme sequences using G2P:
\[
\mathcal{P}^{w}=\{P_1, P_2, \ldots, P_L\}
\]
\State Flatten word-level phonemes into a canonical phoneme sequence:
\[
P=\{p_i\}_{i=1}^{N}
\]
\State Record the ending phoneme index $e_l$ for each word $l$

\State Parse each DA-DPPG token $y_j=\texttt{<p\_v\_d>}$ and discard variant and duration:
\[
Q=\{q_j\}_{j=1}^{M}
\]

\State Initialize dynamic programming table $D \in \mathbb{R}^{(N+1)\times(M+1)}$ with $\infty$
\State Set $D_{0,0}=0$

\For{$i=0$ to $N$}
\For{$j=0$ to $M-1$}
\If{$D_{i,j}=\infty$}
\State continue
\EndIf
    \If{$i>0$}
        \State $c \leftarrow d_{\mathrm{pamc}}(p_i, q_{j+1})$
        \State Update $D_{i,j+1}$ with $D_{i,j}+c$
        \State Store backpointer for extending the current phoneme span
    \EndIf

    \If{$i<N$}
        \State $c \leftarrow d_{\mathrm{pamc}}(p_{i+1}, q_{j+1})$
        \State Update $D_{i+1,j+1}$ with $D_{i,j}+c$
        \State Store backpointer for advancing to the next phoneme
    \EndIf
\EndFor
\EndFor

\State Backtrace from $(N,M)$ to obtain the ending token index $a_i$ for each canonical phoneme $p_i$

\State Initialize token start index $s \leftarrow 1$
\For{each word $l=1$ to $L$}
\State $t \leftarrow a_{e_l}$
\State Assign tokens $\{y_s,\ldots,y_t\}$ to word $l$
\State $\mathcal{A}[l] \leftarrow (s,t)$
\State $s \leftarrow t+1$
\EndFor

\State \Return $\mathcal{A}$
\end{algorithmic}
\end{algorithm}

\section{Additional Phoneme Identity Editing Experimental Results}

\subsection{Experiment}
To evaluate the generality of phoneme identity editing, we construct eight phoneme substitution pairs spanning nasals, liquids, fricatives, stops, and vowels. For each phoneme, we select its canonical DPPG variant, defined as the variant whose corresponding PPG cluster exhibits the highest posterior probability for the target phoneme identity. This selection avoids context-dependent variants associated with coarticulation, allowing the evaluation to focus on phoneme identity. For each pair, we replace the source phoneme token with the target phoneme token while preserving the remaining DPPG tokens unchanged and synthesize the edited speech.
To evaluate phoneme identity editing, the synthesized speech is processed by a pretrained phoneme recognizer~\cite{churchwell2024ppg}, and frame-level phoneme identities are obtained by taking the argmax of the predicted phoneme posteriorgrams (PPGs).
We report three metrics TPD, SPD and OPD as described in the paper. For each edited sample, phoneme identities are analyzed within a local window centered at the edited position.

\subsection{Result and analysis}
As shown in Table~\ref{tab:phoneme_identity_supp}, phoneme identity editing achieves an average TPD of 83.75\% with a low SPD of 7.50\%. This indicates that the source phoneme is usually suppressed and the target phoneme is detected in most cases, demonstrating that DPPG-based explicit token manipulation enables phoneme-level editing within UniSAE.

Meanwhile, the detection rate varies across phoneme pairs. We hypothesize that editing difficulty is influenced by both lexical validity and local phonetic context. When the edited phoneme produces a valid word (e.g., `nine'' $\rightarrow$ `line'' or `tom'' $\rightarrow$ `dom''), the target phoneme is often realized successfully. In contrast, many phoneme substitutions result in non-existent words (e.g., `please'' $\rightarrow$ `blease''), making the edited segment more susceptible to contextual influences from neighboring phonemes and learned speech priors.

For example, the /p/$\rightarrow$/b/ pair exhibits a relatively low TPD. Upon inspection, many failures result in an ambiguous realization perceptually located between /p/, /b/, and silence. Consequently, the phoneme recognizer frequently fails to detect the target phoneme /b/. Interestingly, although the phoneme-level realization is ambiguous, the edited speech often remains intelligible. An ASR system may still recognize the word as ``please'' by leveraging lexical and contextual information. This observation suggests that phoneme-level editing performance depends not only on the edited phoneme pair itself, but also on lexical plausibility and the surrounding phonetic context. A more systematic investigation of these factors is left for future work.

\begin{table}[t]
\centering
\caption{
Phoneme-level editing result across eight phoneme substitution pairs.
}
\label{tab:phoneme_identity_supp}
\small
\begin{tabular}{lccc}
\toprule
Phoneme Edit
& TPD (\%) $\uparrow$
& SPD (\%) $\downarrow$
& OPD (\%) $\downarrow$
\\
\midrule

/n/ $\rightarrow$ /l/
& 100
& 0
& 0
\\

/r/ $\rightarrow$ /l/
& 70
& 20
& 10
\\

/m/ $\rightarrow$ /n/
& 95
& 5
& 0
\\

/s/ $\rightarrow$ /sh/
& 85
& 5
& 10
\\

/t/ $\rightarrow$ /d/
& 100
& 0
& 0
\\

/p/ $\rightarrow$ /b/
& 60
& 10
& 30
\\

/k/ $\rightarrow$ /g/
& 85
& 10
& 5
\\

/aa/ $\rightarrow$ /ey/
& 75
& 10
& 15
\\

\midrule

Average
& 83.75
& 7.50
& 8.75
\\

\bottomrule
\end{tabular}
\end{table}

\end{document}